\documentclass[twocolumn,english,prc,showpacs]{revtex4}
\usepackage[T1]{fontenc}
\usepackage[latin9]{inputenc}
\usepackage{float}
\usepackage{amsmath}
\usepackage{amssymb}
\usepackage{graphicx}
\usepackage{esint}

\makeatletter

\providecommand{\tabularnewline}{\\}

\@ifundefined{textcolor}{}
{%
 \definecolor{BLACK}{gray}{0}
 \definecolor{WHITE}{gray}{1}
 \definecolor{RED}{rgb}{1,0,0}
 \definecolor{GREEN}{rgb}{0,1,0}
 \definecolor{BLUE}{rgb}{0,0,1}
 \definecolor{CYAN}{cmyk}{1,0,0,0}
 \definecolor{MAGENTA}{cmyk}{0,1,0,0}
 \definecolor{YELLOW}{cmyk}{0,0,1,0}
}

\usepackage{bm}
\newcommand{\pr}[1]{{\sc{\lowercase{#1}}}}

\renewcommand{\vec}[1]{\bm{#1}}

\makeatother

\usepackage{babel}
\begin{document}

\title{Low-lying states in near-magic odd-odd nuclei and the effective interaction}

\author{B.~G.~Carlsson$^{1}$ and J.~Toivanen$^{2}$ }

\affiliation{$^{1}$ Division of Mathematical Physics, LTH, Lund University, Post
Office Box 118, S-22100 Lund, Sweden}

\email{gillis.carlsson@matfys.lth.se}

\affiliation{$^{2}$ Department of Physics, University of Jyväskylä, P.O. Box
35 (YFL) FI-40014, Finland}

\date{\today}
\begin{abstract}
The iterative quasi-particle-random-phase approximation (QRPA) method
we previously developed \cite{Toivanen2010,Vesely2012,Carlsson2012}
to accurately calculate properties of individual nuclear states is
extended so that it can be applied for nuclei with odd numbers of
neutrons and protons. The approach is based on the proton-neutron-QRPA
(pnQRPA) and uses an iterative non-hermitian Arnoldi diagonalization
method where the QRPA matrix does not have to be explicitly calculated
and stored. The method is used to calculate excitation energies of
proton-neutron multiplets for several nuclei. The influence of a pairing
interaction in the $T=0$ channel is studied. 
\end{abstract}


\maketitle

\section{Introduction}

While static properties of atomic nuclei are very interesting, much
can be learned by considering dynamical effects such as the linear
response of nuclei when perturbed by external fields. This can be
modeled using the quasiparticle-random-phase approximation (QRPA)
\cite{Suhonen2007} where the external field excites quasiparticle
pairs. In the standard pp-nnQRPA approach the excitations are composed
of sums of two-proton and two-neutron quasiparticle excitations. If
the field is instead allowed to excite proton-neutron quasiparticle
pairs, the corresponding approximation is denoted pnQRPA \cite{Suhonen2007}.
With the pnQRPA formalism one can model nuclear reactions where in
the final state a proton has turned into a neutron or vice versa as
occurs in the $\beta$-decay processes. However, when modeling $\beta$-decay
using nuclear density-functional theory (DFT), the results are sensitive
to the effective isoscalar pairing interaction used in the model.
Therefore, in recent studies of $\beta$-decay, the isoscalar pairing
interaction is often used as a free fitting parameter \cite{Mustonen2013,Engel1999}.
In order to develop better effective interactions one may try to find
an optimal value for the isoscalar pairing strength without directly
fitting it to the $\beta$-decay probabilities. The challenge in this
respect is that this part of the interaction does not play a role
in standard Hartree-Fock-Bogoliubov (HFB) calculations where pairing
between protons and neutrons are not allowed. Therefore whatever value
is employed does not influence HFB calculations for the ground state.

In a series of papers \cite{Vesely2012,Toivanen2010,Carlsson2013,Carlsson2012}
we have developed fast and memory efficient QRPA solvers which can
easily be used for fine tuning model parameters while taking dynamical
effects into account. In this work we extend these methods to the
pnQRPA case and apply the approach to find the strength of the isoscalar
pairing interaction. A value of the strength is found by using the
pnQRPA to calculate the low-lying spectra of odd-odd nuclei. Starting
from a spherical nucleus and exciting a proton-neutron pair, the particles
can couple their angular momenta forming a multiplet of final angular
momentum values. Without any residual interaction the states of the
multiplet become degenerate but in general they will split apart.
The splitting between such multiplet states was very early interpreted
using empirical rules \cite{Nordheim1950} which stated that the nucleons
prefer to align their intrinsic spins in parallel as in the case of
deuterium. Using a delta interaction the gross features of many such
spectra can be reproduced \cite{Schiffer1971}. In the case of one
proton and one neutron in identical orbits the different states of
the multiplet will alternate between $T=0$ and $T=1$ coupling depending
on weather the total angular momentum is even or odd. Therefore the
splitting of the states in the multiplet is directly sensitive to
the magnitude of the $T=0$ pairing interaction.

In this work we consider the available experimental data for multiplets
and calculate the corresponding states using the pnQRPA formalism.
The strength of the $T=0$ pairing interaction is taken as a free
parameters and is tuned in order to reproduce the experimental multiplet
splittings.

This paper is organized as follows: in Sec. II the pnQRPA formalism
is briefly reviewed and specific aspects of our formulation are discussed.
In Sec III the computational cost and accuracy of the method is evaluated.
In Sec. IV we discuss the experimental data. In sec. V the method
is applied to the calculation of multiplet energies in a selection
of odd-odd nuclei. Finally conclusions are given in section VI.

\section{Theoretical model}

In matrix form the QRPA equation \cite{RingSchuck1980,Suhonen2007}
can be expressed as

\begin{equation}
\hbar\omega\left(\begin{array}{c}
X\\
-Y
\end{array}\right)=\left(\begin{array}{cc}
A & B\\
B^{*} & A^{*}
\end{array}\right)\left(\begin{array}{c}
X\\
Y
\end{array}\right),
\end{equation}
 where the $A$ and $B$ matrices have dimensions the size of two-body
matrix elements. In order to avoid constructing and storing the large
QRPA matrix it is useful to write the action of the matrix on the
QRPA vectors in terms of transitional fields \cite{Carlsson2012,Vesely2012}.
This allows the action of the QRPA matrix on a vector to be constructed
in a three step procedure \cite{Carlsson2012,Vesely2012}. 
\begin{enumerate}
\item In the first step transitional densities are built as: 
\begin{align}
\tilde{\rho} & =U\tilde{Z}V^{T}+V^{*}\tilde{Z}'^{\dagger}U^{\dagger}\\
\tilde{\kappa} & =U\tilde{Z}U^{T}+V^{*}\tilde{Z}'^{\dagger}V^{\dagger}\\
\tilde{\kappa}'^{\dagger} & =V\tilde{Z}V^{T}+U^{*}\tilde{Z}'^{\dagger}U^{\dagger}
\end{align}
 where $U$ and $V$ denotes the matrices of the Bogoliubov transformation
\cite{RingSchuck1980}. $\tilde{Z}$ and $\tilde{Z}'^{\dagger}$ are
antisymmetric matrices whose upper triangular parts correspond to
the elements in the $X$ and $Y$ column vectors. 
\item In the second step the transitional fields are built. In the absence
of a density-dependent pairing interaction they take the form: 
\begin{align}
\tilde{h}_{\mu\nu} & =\sum_{\pi\lambda}\left.\frac{\partial h_{\mu\nu}}{\partial\rho_{\pi\lambda}}\right|_{\rho_{\mathrm{gs}}}\tilde{\rho}_{\pi\lambda}=\sum_{\pi\lambda}\tilde{v}_{\mu\lambda\nu\pi}\tilde{\rho}_{\pi\lambda},\\
\tilde{\Delta}_{\mu\nu} & =\frac{1}{2}\sum_{kl}v_{\mu\nu kl}^{\mathrm{pair}}\tilde{\kappa}_{kl},\\
\left(\tilde{\Delta}'^{\dagger}\right){}_{\mu\nu} & =\frac{1}{2}\sum_{kl}v_{\mu\nu kl}^{\mathrm{pair}*}\left(\tilde{\kappa}'^{\dagger}\right){}_{kl}.
\end{align}
 In these expressions, the matrix elements entering the $\tilde{h}$
expression denotes the effective RPA interaction \cite{RingSchuck1980}
while $v_{\mu\nu kl}^{\mathrm{pair}}$ denotes the pairing two-body
matrix elements. In our case the Skyrme interaction is used as a particle-hole
interaction and a separable interaction is used as a pairing interaction.
With these special interactions, standard methods \cite{Carlsson2010p2,Vesely2012}
can be used to construct the fields which means that one can avoid
constructing large matrices of two-body elements. 
\item In the third step, these fields are multiplied with the Bogoliubov
matrices to form the $\tilde{W}$ matrices: 
\begin{eqnarray}
\tilde{W} & = & U^{\dagger}\tilde{h}V^{*}+U^{\dagger}\tilde{\Delta}U^{*}+V^{\dagger}\tilde{\Delta}'^{\dagger}V^{*}-V^{\dagger}\tilde{h}^{T}U^{*}\\
\tilde{W}'^{\dagger} & = & V^{T}\tilde{h}U+V^{T}\tilde{\Delta}V+U^{T}\tilde{\Delta}'^{\dagger}U-U^{T}\tilde{h}^{T}V.\label{eq:W-1-2-1}
\end{eqnarray}

\end{enumerate}
It should be noted that the steps of building the transitional densities
and fields are analogous to the way of building the HFB densities
and fields and can thus can be performed with slight modifications
to an existing HFB code. Once these steps are completed the QRPA equations
can be formulated as \cite{Carlsson2012,Vesely2012}:

\begin{eqnarray}
\hbar\omega\tilde{Z} & = & E\tilde{Z}+\tilde{Z}E+\tilde{W}\\
-\hbar\omega\tilde{Z}'^{\dagger} & = & E\tilde{Z}'^{\dagger}+\tilde{Z}'^{\dagger}E+\tilde{W}'^{\dagger}
\end{eqnarray}
 where $E$ denotes a diagonal matrix composed of the positive eigenvalues
to the HFB equation \cite{Carlsson2012,Vesely2012}.

In our case when the HFB $U$ and $V$ matrices \cite{RingSchuck1980}
do not mix neutrons and protons, these equations can be divided into
two separate uncoupled pieces where one is the standard pp-nnQRPA
equation and the other piece is the pnQRPA equation. To simplify the
notation for the pnQRPA equation we first introduce the matrices:
\begin{equation}
\tilde{Z}=\left(\begin{array}{cc}
z_{1} & z_{2}\\
z_{3} & z_{4}
\end{array}\right),\,\,\tilde{Z}'^{\dagger}=\left(\begin{array}{cc}
\hat{z}_{1} & \hat{z}_{2}\\
\hat{z}_{3} & \hat{z}_{4}
\end{array}\right)
\end{equation}

\begin{equation}
\tilde{W}=\left(\begin{array}{cc}
w_{1} & w_{2}\\
w_{3} & w_{4}
\end{array}\right),\,\,\tilde{W}'^{\dagger}=\left(\begin{array}{cc}
\hat{w}_{1} & \hat{w}_{2}\\
\hat{w}_{3} & \hat{w}_{4}
\end{array}\right).
\end{equation}
 The grouping into four blocks is obtained from ordering the indexes
so that proton states comes before neutron states. Then for example
in the $z_{2}$ and $\hat{z}_{2}$ matrices, the first index refers
to a proton state and the second one to a neutron state. A similar
notation is used for the $\tilde{\kappa},\tilde{\kappa}'^{\dagger}$
and the $\tilde{\rho}$ matrices. In the same way the $U$ ,$V$ and
$E$ matrices also obtain block structures: 
\begin{equation}
U=\left(\begin{array}{cc}
U_{p} & 0\\
0 & U_{n}
\end{array}\right),\,\, V=\left(\begin{array}{cc}
V_{p} & 0\\
0 & V_{n}
\end{array}\right),\,\, E=\left(\begin{array}{cc}
E_{p} & 0\\
0 & E_{n}
\end{array}\right).
\end{equation}
 With this notation the pnQRPA part of the equation can be expressed:
\begin{eqnarray}
\hbar\omega z_{2} & = & E_{p}z_{2}+z_{2}E_{n}+w_{2}\\
-\hbar\omega\hat{z}_{2} & = & E_{p}\hat{z}_{2}+\hat{z}_{2}E_{n}+\hat{w}_{2}.
\end{eqnarray}
 Since the Bogoliubov transformation preserves the proton and neutron
quantum numbers we obtain:

\begin{align}
w_{2} & =U_{p}^{\dagger}h_{2}V_{n}^{*}+U_{p}^{\dagger}\Delta_{2}U_{n}^{*}+V_{p}^{\dagger}\hat{\Delta}_{2}V_{n}^{*}-V_{p}^{\dagger}h_{3}^{T}U_{n}^{*}\\
\hat{w}_{2} & =V_{p}^{T}h_{2}U_{n}+V_{p}^{T}\Delta_{2}V_{n}+U_{p}^{T}\hat{\Delta}_{2}U_{n}-U_{p}^{T}h_{3}^{T}V_{n}
\end{align}
 and

\begin{align}
\left(h_{2}\right)_{pn} & =\sum_{n'p'}\tilde{v}_{pn',np'}\left(\rho_{2}\right)_{p'n'}\\
\left(h_{3}\right)_{np} & =\sum_{n'p'}\tilde{v}_{np',pn'}\left(\rho_{3}\right)_{n'p'}
\end{align}

\begin{align}
\left(\Delta_{2}\right)_{pn} & =\sum_{p'n'}v_{pn,p'n'}^{\mathrm{pair}}\left(\kappa_{2}\right)_{p'n'}\\
\left(\hat{\Delta}_{2}\right)_{pn} & =\sum_{p'n'}v_{pn,p'n'}^{\mathrm{pair}*}\left(\hat{\kappa}_{2}\right)_{p'n'}
\end{align}
 where the $p$ and $p'$ ($n$ and $n'$ ) indexes refer to proton
(neutron) states. The relevant blocks of the transitional densities
are obtained as

\begin{align}
\rho_{2} & =U_{p}z_{2}V_{n}^{T}+V_{p}^{*}\hat{z}_{2}U_{n}^{\dagger}\\
\rho_{3} & =-U_{n}z_{2}^{T}V_{p}^{T}-V_{n}^{*}\hat{z}_{2}^{T}U_{p}^{\dagger}\\
\kappa_{2} & =U_{p}z_{2}U_{n}^{T}+V_{p}^{*}\hat{z}_{2}V_{n}^{\dagger}\\
\hat{\kappa}_{2} & =V_{p}z_{2}V_{n}^{T}+U_{p}^{*}\hat{z}_{2}U_{n}^{\dagger}.
\end{align}
 The equations involve matrix elements of the Skyrme interaction in
the $T=1$ and $T_{Z}=\pm1$ channels that are not active during standard
HFB calculations which do not mix protons and neutrons. The evaluation
of the extra matrix elements thus requires an extension of the usual
method and this extension will be discussed in the next section.

\subsection{Evaluation of fields}

With the Skyrme functional, and in the spin and isospin coupled notation
the potential energy arising from the density-independent two-body
part of the interaction can be expressed \cite{Carlsson2010p2,Carlsson2010}

\begin{align}
\mathcal{E} & =\int\sum_{\alpha\beta t}\mathcal{C}_{\alpha,J}^{t,\beta}\left[\left[\rho_{\beta,J}^{t},\rho_{\alpha,J}^{t}\right]^{0}\right]_{0}d\vec{r}\nonumber \\
 & =\int\sum_{\alpha\beta t}\mathcal{C}_{\alpha,J}^{t,\beta}\sum_{\substack{m_{t}m_{t}'\\
MM'
}
}C_{tm_{t}tm_{t}'}^{00}C_{JM,JM'}^{00}\nonumber \\
 & \times\rho_{\beta,JM}^{tm_{t}}\left(\vec{r}\right)\rho_{\alpha,JM'}^{tm_{t}'}\left(\vec{r}\right)d\vec{r}.\label{eq:EDF}
\end{align}
 In this expression vector (isovector) coupling is denoted by the
square brackets with subscripts (superscripts) giving the value of
the total spin (isospin). The coefficients $\mathcal{C}_{\alpha,J}^{t,\beta}$
denote the coupling constants of the model while e.g. $C_{tm_{t},tm_{t}'}^{00}$
denote Clebsch-Gordan coefficients \cite{Varshalovich1988}. The local
densities entering this expression are defined as

\begin{align}
 & \rho_{\alpha,JM}^{tm_{t}}\left(\vec{r}\right)=\rho_{mI,nL\nu J',JM}^{tm_{t}}\left(\vec{r}\right)\nonumber \\
 & =\left[\hat{D}_{mI},\left.\left[\hat{K}_{nL},\rho_{\nu}^{tm_{t}}\left(\vec{r},\vec{r}'\right)\right]_{J'}\right|_{\vec{r}'=\vec{r}}\right]_{JM}.
\end{align}
 In this expression $\hat{D}_{mIM}$ ($\hat{K}_{nLM}$) denote derivative
operators (relative momentum operators) coupled to spherical tensors
introduced in \cite{Carlsson2008,Carlsson2010p2}. To keep the notation
simple we have introduced the label $\alpha$ that stands for the
set of quantum numbers $\alpha=\left\{ mI,nL\nu J'\right\} $.

The local densities depend on the spin-isospin one-body density defined
as

\begin{align}
\rho_{\nu m_{\nu}}^{tm_{t}}\left(\vec{r},\vec{r}'\right) & =\sum_{\tau\tau'\sigma\sigma'}\rho\left(\vec{r}\tau\sigma,\vec{r}'\tau'\sigma'\right)\nonumber \\
 & \times\left\langle \sigma'\left|\hat{\sigma}_{\nu m_{\nu}}\right|\sigma\right\rangle \left\langle \tau'\left|\hat{\sigma}_{tm_{t}}\right|\tau\right\rangle \nonumber \\
 & =\sum_{\tau\tau'\sigma\sigma'}\sum_{b,b'}\phi_{b}\left(\vec{r}\sigma\right)\rho_{b\tau,b'\tau'}\phi_{b'}^{*}\left(\vec{r}'\sigma'\right)\nonumber \\
 & \times\left\langle \sigma'\left|\hat{\sigma}_{\nu m_{\nu}}\right|\sigma\right\rangle \left\langle \tau'\left|\hat{\sigma}_{tm_{t}}\right|\tau\right\rangle .
\end{align}
 In this expression the label $b$ stands for the quantum numbers
needed to specify the basis states. For example in the case of a harmonic
oscillator basis $b=\left\{ Nljm\right\} $. We assume the same basis
states for neutrons and protons. The quantum number $\tau$ ($\sigma$)
is the isospin (spin) projection. For the Pauli matrices we use the
tensor form of the operators introduced in Eq. 14 and 15 of \cite{Carlsson2010p2}.
Introducing the short-hand notation $\left\langle \sigma'\left|\hat{\sigma}_{\nu m_{\nu}}\right|\sigma\right\rangle =\sigma_{\nu m_{\nu}}^{\sigma'\sigma}$
the explicit relations for the different components of the one-body
density can be written as:

\begin{align}
\rho_{\nu m_{\nu}}^{00}\left(\vec{r},\vec{r}'\right) & =\sum_{\sigma\sigma'bb'}\phi_{b}\left(\vec{r}\sigma\right)\phi_{b'}^{*}\left(\vec{r}'\sigma'\right)\sigma_{\nu m_{\nu}}^{\sigma'\sigma}\nonumber \\
 & \times\left(\rho_{b-\frac{1}{2},b'-\frac{1}{2}}+\rho_{b\frac{1}{2},b'\frac{1}{2}}\right)\label{eq:isoscalar}\\
\rho_{\nu m_{\nu}}^{10}\left(\vec{r},\vec{r}'\right) & =\sum_{\sigma\sigma'bb'}\phi_{b}\left(\vec{r}\sigma\right)\phi_{b'}^{*}\left(\vec{r}'\sigma'\right)\sigma_{\nu m_{\nu}}^{\sigma'\sigma}\nonumber \\
 & \times\left(-i\right)\left(\rho_{b\frac{1}{2},b'\frac{1}{2}}-\rho_{b-\frac{1}{2},b'-\frac{1}{2}}\right)\\
\rho_{\nu m_{\nu}}^{1-1}\left(\vec{r},\vec{r}'\right) & =\sum_{\sigma\sigma'bb'}\phi_{b}\left(\vec{r}\sigma\right)\phi_{b'}^{*}\left(\vec{r}'\sigma'\right)\sigma_{\nu m_{\nu}}^{\sigma'\sigma}\nonumber \\
 & \times\left(-i\right)\left(\sqrt{2}\rho_{b\frac{1}{2},b'-\frac{1}{2}}\right)\\
\rho_{\nu m_{\nu}}^{11}\left(\vec{r},\vec{r}'\right) & =\sum_{\sigma\sigma'bb'}\phi_{b}\left(\vec{r}\sigma\right)\phi_{b'}^{*}\left(\vec{r}'\sigma'\right)\sigma_{\nu m_{\nu}}^{\sigma'\sigma}\nonumber \\
 & \times\left(-i\right)\left(-\sqrt{2}\rho_{b-\frac{1}{2},b'\frac{1}{2}}\right)
\end{align}
 We take protons to have isospin $\tau=-1/2$ so Eq. \ref{eq:isoscalar}
says e.g. that the isoscalar part of the one-body density involves
the sum of the proton and neutron density matrices. Expanding the
isospin coupling in Eq. \ref{eq:EDF} gives

\begin{align}
\mathcal{E} & =\int d\vec{r}\sum_{\alpha\beta,J}\mathcal{C}_{\alpha,J}^{0,\beta}\left[\rho_{\beta,J}^{00},\rho_{\alpha,J}^{00}\right]_{0}\nonumber \\
 & -\frac{\mathcal{C}_{\alpha,J}^{1,\beta}}{\sqrt{3}}\left[\rho_{\beta,J}^{10},\rho_{\alpha,J}^{10}\right]_{0}\nonumber \\
 & +\frac{\mathcal{C}_{\alpha,J}^{1,\beta}}{\sqrt{3}}\left(\left[\rho_{\beta,J}^{11},\rho_{\alpha,J}^{1-1}\right]_{0}+\left[\rho_{\beta,J}^{1-1},\rho_{\alpha,J}^{11}\right]_{0}\right).
\end{align}
 To make the expression more symmetric we introduce new local densities

\begin{align}
\rho_{\alpha,JM}^{+}\left(\vec{r}\right) & =\rho_{\alpha,JM}^{11}\left(\vec{r}\right)+\rho_{\alpha,JM}^{1-1}\left(\vec{r}\right)\\
\rho_{\alpha,JM}^{-}\left(\vec{r}\right) & =\rho_{\alpha,JM}^{11}\left(\vec{r}\right)-\rho_{\alpha,JM}^{1-1}\left(\vec{r}\right)
\end{align}
 which gives

\begin{align}
 & \mathcal{E}=\int d\vec{r}\sum_{\alpha\beta}C_{\alpha,J}^{0,\beta}\left[\rho_{\beta,J}^{00},\rho_{\alpha,J}^{00}\right]_{0}\nonumber \\
 & -\frac{C_{\alpha,J}^{1,\beta}}{\sqrt{3}}\left[\rho_{\beta,J}^{10},\rho_{\alpha,J}^{10}\right]_{0}\nonumber \\
 & +\frac{C_{\alpha,J}^{1,\beta}}{\sqrt{3}}\frac{1}{2}\left(\left[\rho_{\beta,J}^{+},\rho_{\alpha,J}^{+}\right]_{0}-\left[\rho_{\beta,J}^{-},\rho_{\alpha,J}^{-}\right]_{0}\right).
\end{align}
 The new local densities can be considered to be built from the density
matrices

\begin{align}
 & \rho_{\nu m_{\nu}}^{+}\left(\vec{r},\vec{r}'\right)=\sum_{\sigma\sigma'b,b'}\phi_{b}\left(\vec{r}\sigma\right)\phi_{b'}^{*}\left(\vec{r}'\sigma'\right)\sigma_{\nu m_{\nu}}^{\sigma'\sigma}\rho_{b,b'}^{+}\\
 & \rho_{\nu m_{\nu}}^{-}\left(\vec{r},\vec{r}'\right)=\sum_{\sigma\sigma'b,b'}\phi_{b}\left(\vec{r}\sigma\right)\phi_{b'}^{*}\left(\vec{r}'\sigma'\right)\sigma_{\nu m_{\nu}}^{\sigma'\sigma}\rho_{b,b'}^{-},
\end{align}
 where

\begin{eqnarray}
\rho_{b,b'}^{+} & = & i\sqrt{2}\left(\rho_{b-\frac{1}{2},b'\frac{1}{2}}-\rho_{b\frac{1}{2},b'-\frac{1}{2}}\right)\\
\rho_{b,b'}^{-} & = & i\sqrt{2}\left(\rho_{b-\frac{1}{2},b'\frac{1}{2}}+\rho_{b\frac{1}{2},b'-\frac{1}{2}}\right).
\end{eqnarray}
 The new fields needed for pnQRPA have $\tau\ne\tau'$ and can be
written

\begin{align}
h_{b\tau,b'\tau'} & =\frac{\partial\mathcal{E}}{\partial\rho_{b'\tau',b\tau}}\nonumber \\
 & =\frac{\partial\mathcal{E}}{\partial\rho_{b',b}^{+}}\frac{\partial\rho_{b',b}^{+}}{\partial\rho_{b'\tau',b\tau}}+\frac{\partial\mathcal{E}}{\partial\rho_{b',b}^{-}}\frac{\partial\rho_{b',b}^{-}}{\partial\rho_{b'\tau',b\tau}}\nonumber \\
 & =i\sqrt{2}\left(\frac{\partial\mathcal{E}}{\partial\rho_{b',b}^{+}}2\tau+\frac{\partial\mathcal{E}}{\partial\rho_{b',b}^{-}}\right).
\end{align}
 Thus the main task is calculating the fields

\begin{align}
\Gamma_{bb'}^{+} & =\frac{\partial\mathcal{E}}{\partial\rho_{b',b}^{+}}\nonumber \\
 & =\frac{1}{2}\frac{\partial}{\partial\rho_{b'b}^{+}}\int d\vec{r}\sum_{\alpha\beta,J}\frac{C_{\alpha,J}^{1,\beta}}{\sqrt{3}}\left[\rho_{\beta,J}^{+},\rho_{\alpha,J}^{+}\right]_{0}
\end{align}
 and 
\begin{align}
\Gamma_{bb'}^{-} & =\frac{\partial\mathcal{E}}{\partial\rho_{b',b}^{-}}\nonumber \\
 & =\frac{-1}{2}\frac{\partial}{\partial\rho_{b'b}^{-}}\int d\vec{r}\sum_{\alpha\beta,J}\frac{C_{\alpha,J}^{1,\beta}}{\sqrt{3}}\left[\rho_{\beta,J}^{-},\rho_{\alpha,J}^{-}\right]_{0}.
\end{align}
 Except for the constants $\frac{1}{2}$ and $-\frac{1}{2}$ these
fields have the same form as the fields resulting from the isovector
term. Thus the same computer routines can be reused for the calculation
of these new terms.

\subsection{Density-dependent interaction}

Introducing a standard scalar-isocalar density dependence gives the
new term

\begin{equation}
\mathcal{E}^{dd}=\sum_{\alpha\beta}\frac{C_{\alpha,J}^{1,\beta}}{\sqrt{3}}\rho_{0}^{\alpha}\frac{1}{2}\left(\left[\rho_{\beta,J}^{+},\rho_{\alpha,J}^{+}\right]_{0}-\left[\rho_{\beta,J}^{-},\rho_{\alpha,J}^{-}\right]_{0}\right).
\end{equation}
 One realizes that variations of the type

\begin{equation}
\left.\frac{\partial^{2}\mathcal{E}^{dd}}{\partial\rho_{bb'}^{x}\partial\rho_{cc'}^{x}}\right|_{\rho=\rho_{\mathrm{gs}}}
\end{equation}
 where $x=+\,\mathrm{or}\,-$ will give rise to non-zero contributions
and other variations will not give anything. This is because the $\rho_{bb'}^{+}$
and $\rho_{bb'}^{-}$ density matrices are zero in the ground state
when protons and neutrons are uncorrelated. Therefor one only obtains
contributions to the fields

\begin{equation}
\tilde{\Gamma}_{ij}=\sum_{kl}\tilde{v}_{iljk}\tilde{\rho}_{kl},
\end{equation}
 where one of the indexes $i,j$ refers to a proton and the other
one to a neutron. This means that we must consider the matrix elements
$\tilde{v}_{pn'np'}$ and $\tilde{v}_{np'pn'}$ since the other combinations
where the first two indexes refer to the same particle species are
forbidden by charge conservation. The first of these matrix elements
can be expressed \cite{RingSchuck1980}

\begin{eqnarray}
 &  & \tilde{v}_{pn'np'}=\left.\frac{\partial E_{HF}}{\partial\rho_{np}\partial\rho_{p'n'}}\right|_{\rho_{\mathrm{gs}}}\nonumber \\
 & = & \left.\bar{v}_{n'pp'n}\left[\rho\right]+\sum_{jl}\rho_{lj}\left(\frac{\partial\bar{v}_{n'jp'l}\left[\rho\right]}{\partial\rho_{np}}+\frac{\partial\bar{v}_{pjnl}\left[\rho\right]}{\partial\rho_{p'n'}}\right)\right|_{\rho_{\mathrm{gs}}}\nonumber \\
 & + & \left.\frac{1}{2}\sum_{ijkl}\rho_{ki}\frac{\partial\bar{v}_{ijkl}\left[\rho\right]}{\partial\rho_{p'n'}\partial\rho_{np}}\rho_{lj}\right|_{\rho_{\mathrm{gs}}}\nonumber \\
 & = & \bar{v}_{n'pp'n}\left[\rho_{\mathrm{gs}}\right].
\end{eqnarray}
 The last line follows since the density-dependence is explicitly
with respect to the isoscalar density so the variations of the matrix
elements with respect to the mixed proton-neutron densities become
zero. For the $\tilde{v}_{np'pn'}$ combination it works in the same
way. Thus with the standard isoscalar $\rho_{0}^{\alpha}$ density-dependence
there are no additional rearrangement terms appearing in the pnQRPA.

\subsection{Pairing interaction}

For the pairing interaction we adopt a form:

\begin{align}
V\left(\vec{r}_{1},\vec{r}_{2},\vec{r}'_{1},\vec{r}'_{2}\right) & =\delta\left(\vec{R}-\vec{R}'\right)P\left(r\right)P\left(r'\right)\nonumber \\
 & \times\left[G_{1}\hat{\Pi}_{s=0}+G_{0}\hat{\Pi}_{S=1,T=0}\right]
\end{align}
 where

\begin{align}
P\left(r\right) & =\frac{1}{\left(4\pi a^{2}\right)^{3/2}}e^{-\vec{r}^{2}/\left(4a^{2}\right)}\\
\hat{\Pi}_{s=0} & =\frac{1}{2}\left(1-P^{\sigma}\right)\\
\hat{\Pi}_{S=1,T=0} & =\frac{1}{4}\left(1+P^{\sigma}\right)\left(1-P^{\tau}\right).
\end{align}
 Since this interaction has a finite range it leads to convergent
results and no energy cut-off is needed for the pairing space. The
separable structure of the interaction allows an efficient evaluation
of the two-body matrix elements \cite{Vesely2012}. The isovector
part of this interaction was first considered in \cite{Duguet2004}
to parameterize the bare low-momentum potential in the $^{1}S_{0}$
channel. Here the parameterization is straightforwardly extended to
the $T=0$ channel assuming the same radial dependence.

\section{Accuracy and convergence}

In order to find the eigenvalues of the large pnQRPA matrix we use
the Implicitly Restarted Arnoldi method (IRA) \cite{arnoldi-ira,arnoldi-saad}.
With this approach the pnQRPA matrix never has to be built, it is
sufficient to be able to calculate the results of the matrix acting
on an arbitrary vector which can be done as outlined in the previous
section. The method is implemented in an updated version of the \pr{HOSPHE}
(v1.02) \cite{Carlsson2010p2} code.

An example of the calculations is shown in 
\begin{figure}[t]
\includegraphics[bb=0bp 0bp 561bp 584bp,clip,width=0.9\columnwidth]{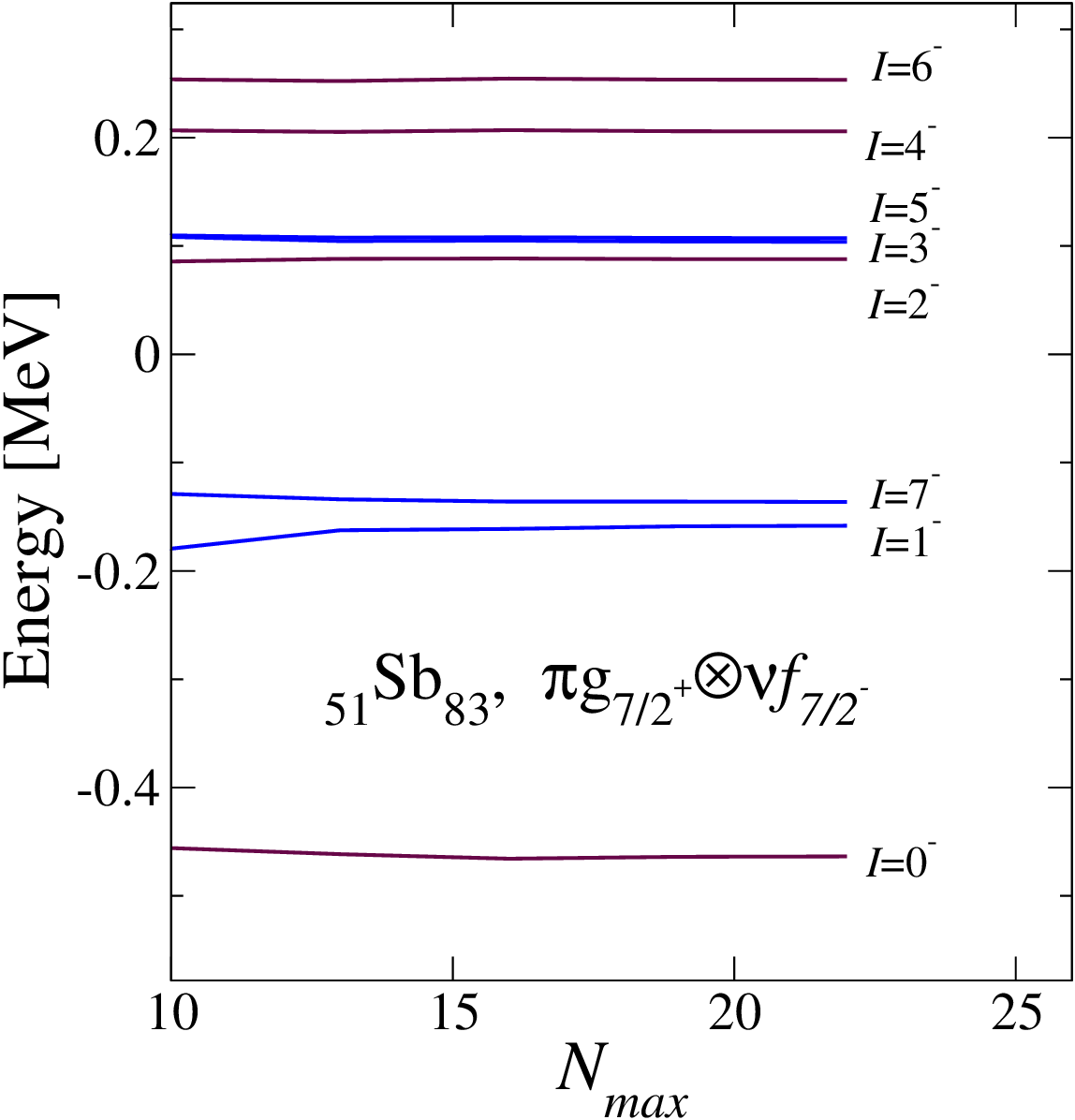}\caption{(Color online) Convergence of the relative energies of the low-lying
multiplet in $_{51}\mathrm{Sb}_{83}$ as a function of the maximum
oscillator shell $N_{max}$ included in the basis. The energies are
drawn relative to the average energy of the multiplet.\label{fig:convergence}
The SKX Skyrme interaction \cite{Brown1998} was used with pairing
parameters $\left(G_{1},G_{0}\right)=\left(545,763\right)$ MeVfm$^{3}$
($G_{0}/G_{1}$=1.4).}
\end{figure}

Fig.~\ref{fig:convergence}. In this figure we have selected the
lowest multiplet states in the nucleus $_{51}$Sb$_{83}$ that has
a proton-neutron pair outside closed shells. The proton neutron pair
is assumed to be in a $\pi g_{7/2}\otimes\nu f_{7/2}$ configuration
and the corresponding excitations are extracted from the code requesting
the states where this configuration has the largest amplitude. The
longest time is spent on calculating the highest angular momentum
states. For the $I=7^{-}$ calculation with $N_{max}=16$ it takes
about 6.5 min on a standard desktop computer (Intel Core i7-2600K,
3.4GHz). The time depends on the requested accuracy as well as the
number of requested converged excitations. In this case the 15 lowest
positive energy pnQRPA excitations was requested and set to be converged
with a tolerance parameter \cite{ARPACK} of $10^{-6}$. As seen from
Fig.~\ref{fig:convergence}, the relative energies of the multiplet
states converge rapidly with increasing number of oscillator shells.
$N_{max}=16$ appears to give a sufficient accuracy and will therefore
be used in the following.

\section{Selection of experimental data}

Starting from double-magic spherical nuclei we consider neighboring
nuclei with an excited proton-neutron pair of particles or holes.
The proton-neutron pair can couple to different total angular momentum
values forming a multiplet of states. In order to identify the states
we start by considering the experimental ground states of the odd
nuclei surrounding the double magic one. From the ground state spins
of the odd nuclei, $j_{p}$ and $j_{n}$ we can identify the corresponding
configurations by comparing with a Nilsson diagram. Then the largest
components in the lowest states of the odd-odd nuclei are assumed
to result from the coupling of these states. In the case of $N=Z$
nuclei the isospin of the states is sometimes experimentally determined.
In these cases we apply the additional condition that for protons
and neutrons in identical orbits, even (odd)$J$ must be combined
with $T=1$ (0) in order to make the wave function anti symmetric
\cite{Talmi}. Thus for the $N=Z$ nuclei we make use of this relation
and select the lowest experimental states that have isospin values
consistent with those of our assumed configurations. 
\begin{table}[H]
\begin{tabular}{ccccc}
\hline 
Nucleus  & Configuration  & $I^{\pi}$  & $E_{exp}$  & Remark\tabularnewline
\hline 
$_{\,9}^{18}\mathrm{F}_{9}$  & $\pi d_{5/2^{+}}\otimes\nu d_{5/2^{+}}$  & $0^{+}$  & 1.04155  & \tabularnewline
 &  & $1^{+}$  & 0  & (a)\tabularnewline
 &  & $2^{+}$  & 3.06184  & \tabularnewline
 &  & $3^{+}$  & 0.93720  & (a)\tabularnewline
 &  & $4^{+}$  & 4.65200  & \tabularnewline
 &  & $5^{+}$  & 1.12136  & \tabularnewline
$_{21}^{42}\mathrm{Sc}_{21}$  & $\pi f_{7/2^{-}}\otimes\nu f_{7/2^{-}}$  & $0^{+}$  & 0  & \tabularnewline
 &  & $1^{+}$  & 0.611051  & \tabularnewline
 &  & $2^{+}$  & 1.58631  & \tabularnewline
 &  & $\left(3^{+}\right)$  & 1.49043  & \tabularnewline
 &  & $4^{+}$  &  & \tabularnewline
 &  & $\left(5^{+}\right)$  & 1.51010  & \tabularnewline
 &  & $6^{+}$  &  & \tabularnewline
 &  & $\left(7^{+}\right)$  & 0.61628  & \tabularnewline
$_{21}^{50}\mathrm{Sc}_{29}$  & $\pi f_{7/2^{-}}\otimes\nu p_{3/2^{-}}$  & $2^{+},3^{+}$  & 0.256895  & \tabularnewline
 &  & $\left(3^{+}\right)$  & 0.328447  & \tabularnewline
 &  & $\left(4^{+}\right)$  & 0.757000  & \tabularnewline
 &  & $5^{+}$  & 0  & \tabularnewline
$_{29}^{58}\mathrm{Cu}_{29}$  & $\pi p_{3/2^{-}}\otimes\nu p_{3/2^{-}}$  & $0^{+}$  & 0.202990  & \tabularnewline
 &  & $1^{+}$  & 0  & \tabularnewline
 &  & $2^{+}$  & 1.6525  & \tabularnewline
 &  & $\left(3^{+}\right)$  & 0.443640  & \tabularnewline
$_{\,51}^{134}\mathrm{Sb}_{83}$  & $\pi g_{7/2^{+}}\otimes\nu f_{7/2^{-}}$  & $\left(0^{-}\right)$  & 0  & \tabularnewline
 &  & $\left(1^{-}\right)$  & 0.0130  & \tabularnewline
 &  & $\left(2^{-}\right)$  & 0.3311  & \tabularnewline
 &  & $\left(3^{-}\right)$  & 0.3840  & \tabularnewline
 &  & $\left(4^{-}\right)$  & 0.5550  & \tabularnewline
 &  & $\left(5^{-}\right)$  & 0.441  & \tabularnewline
 &  & $\left(6^{-}\right)$  & 0.617  & \tabularnewline
 &  & $\left(7^{-}\right)$  & 0.279  & \tabularnewline
 & $\pi d_{5/2^{+}}\otimes\nu f_{7/2^{-}}$  & $\left(1^{-}\right)$  & 0.8850  & \tabularnewline
 &  & $\left(2^{-}\right)$  & 0.9350  & \tabularnewline
 &  & $3^{-}$  &  & \tabularnewline
 &  & $4^{-}$  &  & \tabularnewline
 &  & $5^{-}$  &  & \tabularnewline
 &  & $6^{-}$  &  & \tabularnewline
 &  &  &  & \tabularnewline
$_{\,83}^{210}\mathrm{Bi}_{127}$  & see caption  &  &  & \tabularnewline
\end{tabular}

\caption{Experimental data \cite{ENSDF} for nuclei with a proton-neutron pair
outside closed shells. The experimental states considered are listed
along with their assumed largest configurations. For $_{\,83}^{210}\mathrm{Bi}_{127}$
we have adopted the first seven multiplets shown in Table III of Ref.
\cite{Bi210} along with the suggested 58 corresponding experimental
energies. All energies are in MeV.\label{tab:Experimental-data1}}
\end{table}

In this way tables \ref{tab:Experimental-data1} and \ref{tab:Experimental2}
are constructed. Table \ref{tab:Experimental-data1} contains data
for particle states and table \ref{tab:Experimental2} contains data
for hole states. In addition to these tables, Ref. \cite{Bi210} contains
a table of 13 identified experimental multiplets in $_{\,83}^{210}\mathrm{Bi}_{127}$.
As part of the data set we adopt the first 7 multiplets shown in Table
III of Ref.~\cite{Bi210}. 
\begin{table}[H]
\begin{tabular}{ccccc}
\hline 
Nucleus  & Configuration  & $I^{\pi}$  & $E_{exp}$  & Remark\tabularnewline
\hline 
$_{\,7}^{14}\mathrm{N}_{7}$  & $\pi p_{1/2^{-}}\otimes\nu p_{1/2^{-}}$  & $0^{+}$  & 2.312798  & \tabularnewline
 &  & $1^{+}$  & 0  & \tabularnewline
$_{19}^{38}\mathrm{K}_{19}$  & $\pi d_{3/2^{+}}\otimes\nu d_{3/2^{+}}$  & $0^{+}$  & 0.1304  & \tabularnewline
 &  & $1^{+}$  & 1.698  & \tabularnewline
 &  & $2^{+}$  & 2.40107  & \tabularnewline
 &  & $3^{+}$  & 0  & \tabularnewline
$_{19}^{46}\mathrm{K}_{27}$  & $\pi d_{3/2^{+}}\otimes\nu f_{7/2^{-}}$  & $\left(2^{-}\right)$  & 0  & \tabularnewline
 &  & $3^{-}$  & 0.5874  & (a)\tabularnewline
 &  & $\left(4^{-}\right)$  & 0.6909  & (a)\tabularnewline
 &  & $5^{-}$  & 0.8855  & \tabularnewline
$_{27}^{54}\mathrm{Co}_{27}$  & $\pi f_{7/2^{-}}\otimes\nu f_{7/2^{-}}$  & $0^{+}$  & 0  & \tabularnewline
 &  & $1^{+}$  & 0.93690  & \tabularnewline
 &  & $2^{+}$  & 1.44566  & \tabularnewline
 &  & $3^{+}$  & 1.82149  & \tabularnewline
 &  & $4^{+}$  & 2.65197  & (b)\tabularnewline
 &  & $\left(5^{+}\right)$  & 1.8870  & \tabularnewline
 &  & $\left(6^{+}\right)$  & 2.979  & (b)\tabularnewline
 &  & $7^{+}$  & 0.1970  & \tabularnewline
$_{\,49}^{130}\mathrm{In}_{81}$  & $\pi g_{9/2^{+}}\otimes\nu h_{11/2^{-}}$  & $\left(1^{-}\right)$  & 0.0000  & \tabularnewline
 &  & $2^{-}$  &  & \tabularnewline
 &  & $3^{-}$  &  & \tabularnewline
 &  & $4^{-}$  &  & \tabularnewline
 &  & $5^{-}$  &  & \tabularnewline
 &  & $6^{-}$  &  & \tabularnewline
 &  & $7^{-}$  &  & \tabularnewline
 &  & $8^{-}$  &  & \tabularnewline
 &  & $9^{-}$  &  & \tabularnewline
 &  & $\left(10^{-}\right)$  & 0.0500  & \tabularnewline
\end{tabular}

\caption{Same as Table \ref{tab:Experimental-data1} but for nuclei with a
proton-neutron hole-pair outside closed shells. Levels marked with
(a) may belong to the multiplet $\pi s_{1/2^{+}}\otimes\nu f_{7/2^{-}}$.
\label{tab:Experimental2}}
\end{table}

In some cases it is possible to compare our assumed assignments for
the largest wave function configurations with previous shell-model
calculations. In the case of $^{18}$F shell-model calculations \cite{KuoBrown2}
confirm our assumptions about the largest amplitude configurations
except for the $1^{+}$ and $3^{+}$ states where the largest components
are suggested to be $\pi d_{5/2}\nu d_{3/2}$ and $\pi d_{5/2}\nu s_{1/2}$
configurations. These two states marked with (a) in Tab. \ref{tab:Experimental-data1}
are thus excluded from the data set. For $^{42}$Sc our assumption
about the largest amplitude configurations is confirmed by shell-model
calculations \cite{KuoBrown}. In $^{50}$Sc there are two possible
spin assignments for the second state of the multiplet. Previous comparisons
with shell-model calculations \cite{Moazed1969} suggest the $2^{+}$
interpretation we adopt here as well. In the case of $^{134}$Sb,
the spin values are shown in parenthesis indicating that they are
not directly measured but comparisons with shell model calculations
\cite{Coraggio2006} support the experimental spin assignments given
for the two observed multiplets in this nucleus.

In the case of hole states shown in Tab. \ref{tab:Experimental2}
and for $^{38}$K we have chosen the second observed $1^{+}$ state
that experiments suggests to be the one with largest $d_{3/2}$ components
\cite{Fenton}. For $^{46}$K we have excluded the $3^{-}$ and the
$4^{-}$ states marked with (a) in Tab. \ref{tab:Experimental2} from
the data set. That is because these states may also arise from a possible
$\left[\pi s_{1/2^{+}},\nu f_{7/2^{-}}\right]_{3^{-},4^{-}}$ coupling
or may be a mixture of both of these multiplets. Although the pnQRPA
takes this mixing into account we prefer to have as clean data as
possible. In the case of $^{54}$Co we adopt the $4^{+}$ and $6^{+}$
states marked with (b) in Tab. \ref{tab:Experimental2} although there
are lower states with tentative spin assignments that could belong
to the multiplet. The decay patterns and comparisons with shell model
calculations suggest the present interpretation \cite{Rudolph2010}.
With these selections we end up with a data set consisting of a total
of 104 states.

\section{Determination of the isoscalar pairing strength}

In this article the values of the isovector pairing strengths and
range parameter ($a=0.660$ fm) are considered to be fixed from values
used in our previous study \cite{Carlsson2012}. In Ref. \cite{Carlsson2012}
different strength was used for neutrons and protons but in this work
we assume an isospin symmetric $T=1$ interaction with a strength
given by the average of the proton and neutron values taken from \cite{Carlsson2012}.
In general the Coulomb interaction will introduce isospin breaking
leading to different pairing strengths for protons and neutrons. In
a complete approach one should thus also consider the Coulomb contribution
to the pairing interaction. However, inclusion of Coulomb is problematic
since approximate treatments may give rise to divergences, see e.g.~\cite{Carlsson2013},
and exact treatments becomes time-consuming. Since the objective of
this work is to determine a first value of the $T=0$ pairing strength
that can be used in pnQRPA calculations for $\beta$-decay, we have
opted to start by investigating the simpler isospin invariant form.

When comparing experimental and theoretical states one should note
that in the pnQRPA formalism the resulting excitations have preserved
total angular momentum and parity but are in general composed of a
mixture of 'pure' multiplet configurations such as those shown in
Tabs. \ref{tab:Experimental-data1} and \ref{tab:Experimental2}.
Thus in order to select the states that should be compared with data
we extract the theoretical states that has the postulated experimental
configurations as the largest amplitudes. In case there are two such
theoretical states the one lowest in energy is selected.

\subsection{Full fits}

\begin{figure}[t]
\includegraphics[clip,width=0.9\columnwidth]{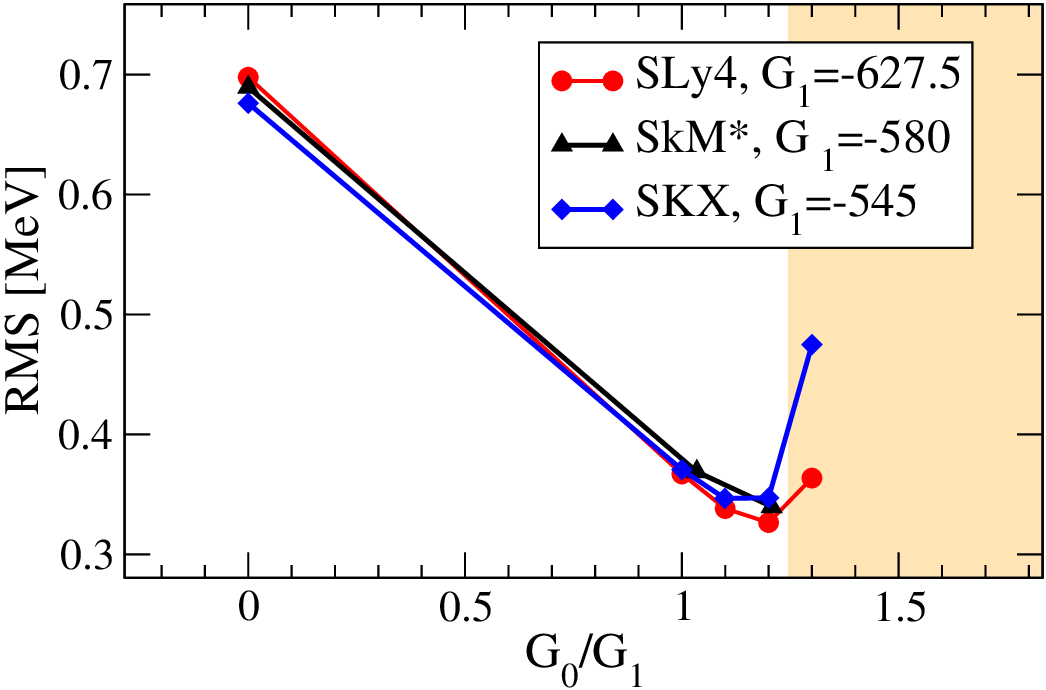}

\caption{\label{fig:Full-RMS}(Color online) RMS as a function of the $G_{0}/G_{1}$
ratio. All 104 experimental states was used for the comparison.}
\end{figure}

In the case of a multiplet where all experimental states are not measured
or some states are excluded on the basis of being uncertain we define
the average energy of the multiplet as the average of the remaining
experimental states. The average of the same theoretical states are
then used to define the average theoretical energy of the multiplet.
Since in general Skyrme interactions will produce errors of $\sim1.4$
MeV for single-particle energies \cite{Kortelainen2008} we do not
compare the average energies with experiment. Instead the experimental
and theoretical relative energies within the multiplet are compared
and the RMS is taken as the difference between experimental and theoretical
relative energies.

Fig.~\ref{fig:Full-RMS} shows the RMS as a function of the isoscalar
pairing strength. The data set involved all 104 states and as seen
in the figure the description of data becomes better as the strength
is increased.

The curves in Fig.~\ref{fig:Full-RMS} are drawn until imaginary
eigenvalues starts to appear in the pnQRPA calculations. For each
interaction, starting from the last point on the curves and increasing
the $G_{0}/G_{1}$ ratio by 10 \% leads to the appearance of such
points. For all nuclei it seems that as the $T=0$ strength reaches
a value $G_{0}\gtrsim1.2G_{1}$ the pnQRPA starts to become unstable
for $N=Z$ nuclei. This may indicate that the ground state is not
a stationary point with respect to proton-neutron correlations and
may thus go through a transition into an isoscalar proton-neutron
pairing condensate (see e.g. the discussion in Ref. \cite{Satula2000}).

Both the SKX \cite{Brown1998} and the SLy4 \cite{Chabanat1998} interactions
show minima at $G_{0}/G_{1}=1.2$ and in both cases the increase in
RMS when going to $G_{0}/G_{1}=1.3$ can be traced to two $N=Z$ nuclei
($^{42}$Sc and $^{38}$K) whose errors increase substantially while
the RMS for most of the remaining nuclei actually decreases. Further
increasing the $T=0$ pairing strength to $G_{0}/G_{1}=1.4$ leads
to imaginary eigenvalues appearing in the $J=1^{+}$ channel for the
same nuclei. At $G_{0}/G_{1}=1.3$ and for the SKX interaction the
lowest energy excitation in this channel is at 0.18 MeV while it is
at 1.9 MeV with SLy4 indicating that the SLy4 minimum is more reliable
while the last SKX point is likely too close to instability to be
reliable.

The results for the SkM{*} interaction \cite{Bartel1982} follows
the other ones but reaches the unstable point before any tendency
for a minimum is displayed.

In general the multiplet splitting is larger in the light nuclei
and they therefor get more important when tuning the strength. In
order to remove this dependence one can divide the energies by $41A^{-1/3}$
to obtain oscillator units \cite{Nilsson1995} which removes the average
energy dependence arising from the different stiffness of the nuclear
potential for light and heavy nuclei. If the RMS is instead calculated
in oscillator units the minimum obtained for SLy4 still occurs for
the same interaction strength.

\subsection{Fits with a reduced data set}

In order to be able to test a larger range of interaction strengths
the $Z=N$ nuclei are excluded from the fits leaving a total of 76
states. The result of this calculation is shown in 
\begin{figure}[t]
\includegraphics[width=0.9\columnwidth]{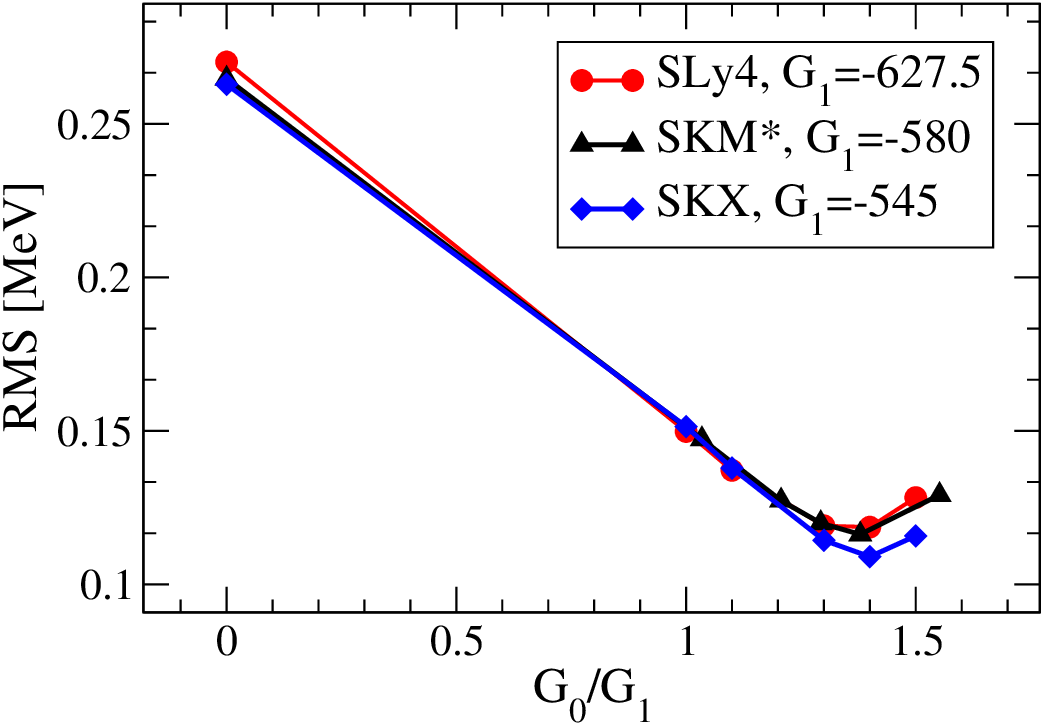}\caption{\label{fig:RMS-no_ZN}(Color online) RMS as a function of the $G_{0}/G_{1}$
ratio. The comparison is performed using the 76 states remaining when
$N=Z$ nuclei are removed from the data set.}
\end{figure}

Fig.~\ref{fig:RMS-no_ZN}. As seen in this figure, when the $N=Z$
nuclei are removed all the interactions produce minima when $G_{0}\simeq1.4G_{1}$.
It is interesting to note that the obtained values are in good agreement
with the ratio of isovector to isoscalar pairing of 1.3 that was found
in Ref. \cite{Satula2000} in order to describe the Wigner energy
as a binding energy gain caused by $T=0$ pairing in the BCSLN model.

\subsection{Results for multiplets}

The results from the optimal fit obtained with the SKX interaction
and $G_{0}=1.4G_{1}$ are shown for the largest multiplets in 
\begin{figure}[t]
\includegraphics[bb=0bp 45bp 359bp 609bp,clip,width=0.9\columnwidth]{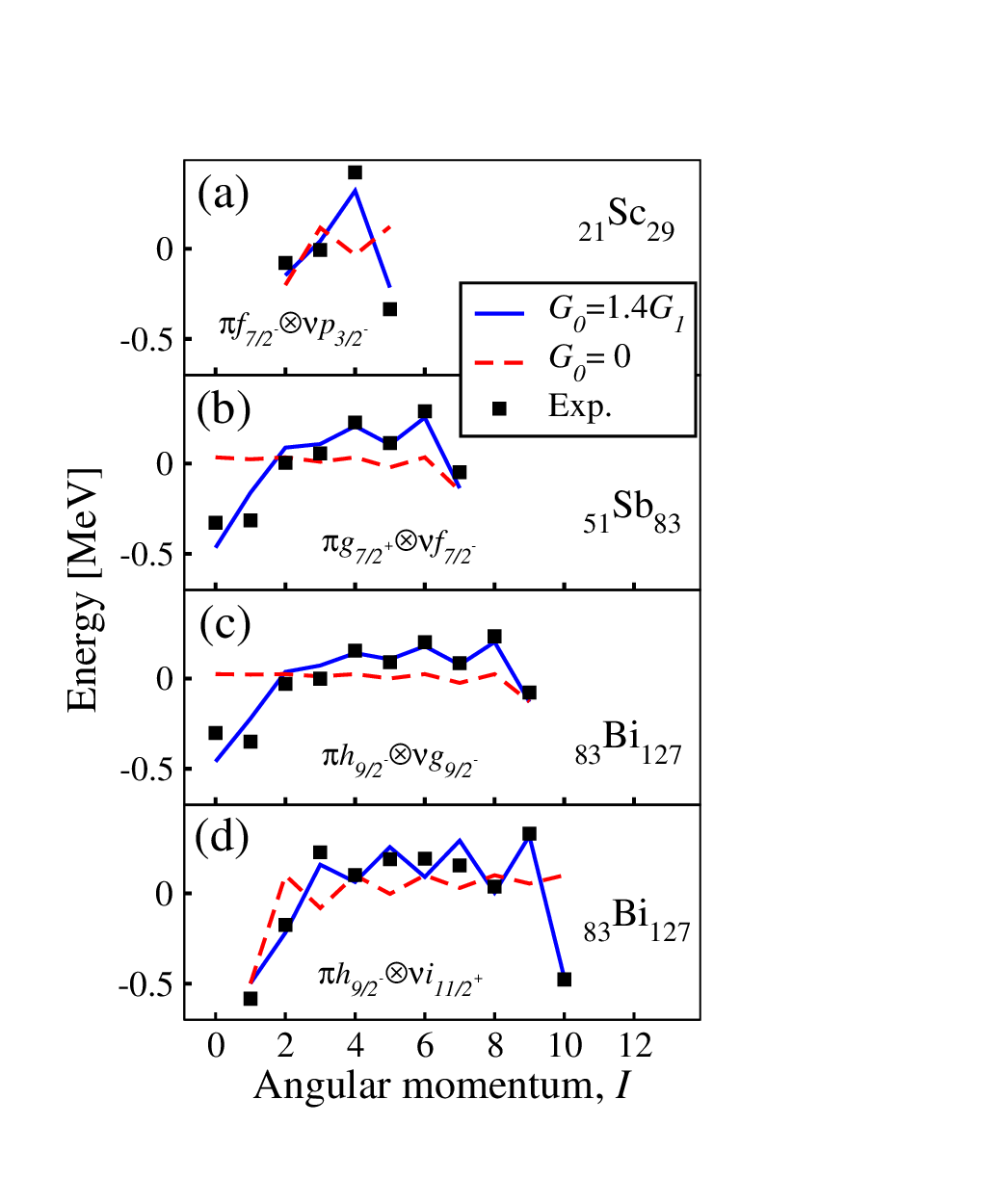}\caption{\label{fig:SKX-mult1}(Color online) Relative energies of the largest
$N\ne Z$ multiplets calculated using the SKX interaction and compared
with experiment. }
\end{figure}

\begin{figure}[t]
\includegraphics[bb=0bp 0bp 349bp 706bp,clip,width=0.9\columnwidth]{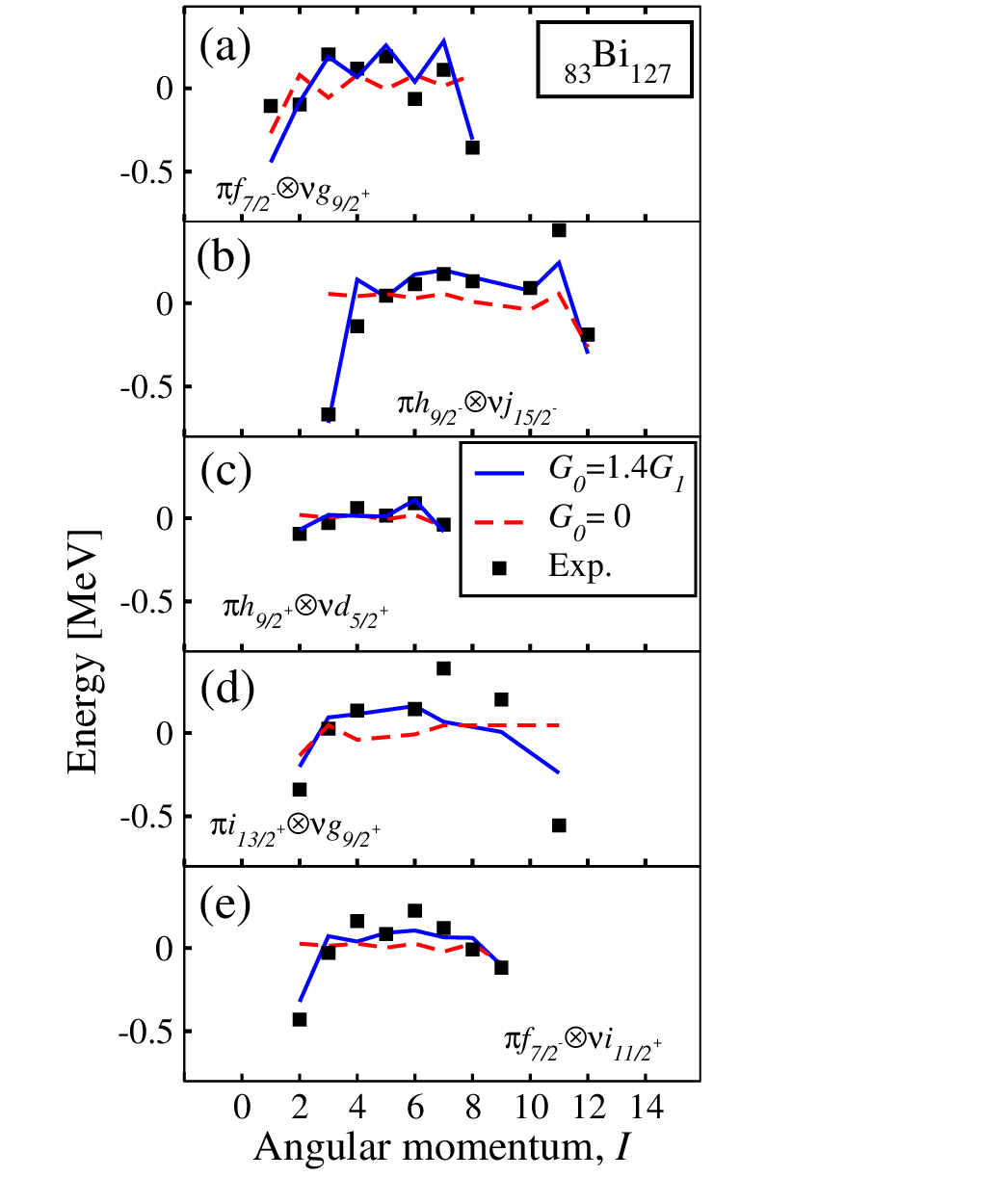}\caption{\label{fig:SKX-mult2}(Color online) Same as Fig.~\ref{fig:SKX-mult1}
for additional multiplets.}
\end{figure}

figures \ref{fig:SKX-mult1} and \ref{fig:SKX-mult2}. If it was not
for the the isoscalar and isovector pairing interactions the resulting
theoretical curves would become constant with the value 0. Thus no
splitting of the multiplets would be predicted. Including a $T=1$
pairing interaction results in the dashed curves in Figs.~\ref{fig:SKX-mult1}
and \ref{fig:SKX-mult2}. These curves obtain some staggering that
makes them agree better with experiment.

With the $T=0$ pairing interaction added, shown with full drawn curves
in figures \ref{fig:SKX-mult1} and \ref{fig:SKX-mult2}, the description
shows a considerable improvement and the theoretical multiplet splittings
are in good agreement with experiment.

For the $_{21}\mathrm{Sc}_{29}$ nucleus the lowest multiplet is expected
to result from a $\pi\left(f_{7/2}\right)^{1}\nu\left(p_{3/2}\right)^{1}$
configuration which can couple to $I=2^{+}-5^{+}$\cite{Moazed1969}.
As seen from panel (a) in Fig.~\ref{fig:SKX-mult1}, the ordering
of the states is correctly described and the relative energies compare
well with experiment. The lowest multiplet in $_{51}\mathrm{Sb}_{83}$
was previously described in the shell-model approach using experimental
single-particle levels and an effective interaction derived from the
CD-Bonn $NN$ potential \cite{Coraggio2006}. The biggest discrepancy
was obtained for the $7^{-}$ state which was predicted about 130
keV above the experimental state. In our case, the $7^{-}$ state
shown in panel (b) of Fig.~\ref{fig:SKX-mult2} is instead predicted
about 130 keV below the experimental state. However it should also
be noted that in Ref. \cite{Coraggio2006} the energies are normalized
to the lowest state in the multiplet while in this work we normalize
to the average multiplet energy. The most striking difference between
the calculations is that while we overpredict the relative energies
of the $0^{-}$ and $1^{-}$ states, the shell model calculation gave
almost the correct energy splitting between these states. The same
discrepancy is also seen in $_{83}\mathrm{Bi}_{127}$ (see panel (c)
of Fig.~\ref{fig:SKX-mult1}) where the $0^{-}$ state comes out
lowest in our calculation while the experimental ground state is $1^{-}$.
Both shell model calculations based on realistic interactions \cite{Coraggio2007}
and phenomenological forces that include non-central tensor and spin-orbit
terms can give the correct ordering \cite{Bi210}. For example in
Ref. \cite{Bi210} a phenomenological force with 8 free parameters
was fitted to multiplet data in $_{83}\mathrm{Bi}_{127}$ which lead
to a good reproduction of the spectra. This suggests that a more complicated
interaction could give a better description of the data but rather
than introducing additional parameters it seems more interesting to
attempt to constrain the $T=0$ effective force starting from bare
interactions as done for the $T=1$ part in Ref. \cite{Duguet2004}.
However this work is left as a future exercise.

In Fig.~\ref{fig:SKX-mult2} the description of the higher lying
multiplets in $_{83}\mathrm{Bi}_{127}$ is shown. A few more multiplets
have been identified \cite{Bi210} but here we have restricted us
to those with the lowest excitation energies since the mixing with
configurations that are outside the scope of the pnQRPA description
are expected to increase with increasing excitation energy.

\section{Summary and conclusions}

An iterative method for the solution of the pnQRPA equations that
avoids the construction of the large pnQRPA matrix was introduced
and employed for the calculation of low-lying states. The method uses
the Implicitly Restarted Arnoldi approach for the solution of the
non-hermitian eigenvalue problem. In this approach, only the action
of the matrix on a Ritz vector is needed and this can be expressed
in terms of effective fields generated by transitional densities.
The numerical tests shows that the method is both fast and reliable.
When generalizing the method to the pnQRPA case additional fields
in the particle-hole channel which are not active in standard HFB
calculations must be taken into account. The expressions for the new
fields follow straightforwardly from the requirement that the nuclear
interaction is invariant with respect to rotations in isospin space
and we demonstrated how they may be calculated analogously to the
standard isovector fields.

The excitations in the pnQRPA are proton and neutron quasiparticle
pairs and the results become sensitive to weather these pairs like
to pair up with their spins in parallel or anti-parallel. This feature
is determined by the relative strengths of the $T=1$ and $T=0$ components
of the pairing interaction. It should be noted that in recent fits
of Skyrme interactions \cite{Goriely2013,Kortelainen2013} the $T=0$
pairing channel is not probed at all since proton-neutron pairing
is generally neglected in the models. However for descriptions of
$\beta$-decay \cite{Mustonen2013,Engel1999} and neutrinos that scatter
on nuclei \cite{Almosly} the $T=0$ pairing channel has a large influence
on the results.

In this work we considered a simple isospin invariant parameterization
of the $T=1$ and $T=0$ pairing interactions and determined the $T=0$
pairing strength from multiplet data. The comparison with experimental
data suggests that the effective pairing interaction in the $T=0$
channel should be roughly 40 \% stronger than the $T=1$ pairing interaction.
It is interesting that these values are in agreement with previous
estimates of a 30 \% stronger $T=0$ channel obtained from assuming
the Wigner energy arises from proton-neutron pairing \cite{Satula2000}.
The collapse of the pnQRPA obtained for some of the $N=Z$ nuclei
further corroborates this view and may be indicative of a phase transition
to a $T=0$ pairing condensate.

It should also be noticed that in a recent study \cite{Bentley2013}
a reasonable agreement with experiment was obtained for Wigner energies
using an effective Hamiltonian without any isoscalar pairing.
In fact the authors find that, with their model, ratios of isoscalar
to isovector pairing larger than 0.8 does not lead to the correct
mass differences. 

It is not straightforward to compare the different studies since different
many-body approaches and different interactions are generally employed. 
It would therefore be interesting to see if
it is possible to find a common effective $T=0$ interaction compatible
with all the different types of experimental data such as multiplet 
energies, $\beta$-decay probabilities, ground state energies etc. 

Although the many-body approach of this work is more advanced, the
structure of the pairing interactions assumed here is certainly simpler
than those of some previous studies \cite{Schiffer76,Bi210}. However
since the main features of the data can be described with the simple
form used in this work, it may be taken as a first approximation to
be used in future pnQRPA studies. In the long run the goal would be
to determine better effective interactions and preferably such interactions
that have the same form in both in the particle-hole and the paring
channels. For such studies low-lying states in odd-odd nuclei can
provide important constraints on the effective interactions.

\section*{acknowledgments}

B.G. Carlsson acknowledges D. Rudolph for discussions concerning the
experimental data and the Royal Physiographic Society in Lund for
providing funding for the computers on which the calculations were
performed. We also thank R. Bengtsson for valuable comments on the
manuscript. This work was supported in part by the Academy of Finland
and University of Jyväskylä within the FIDIPRO programme. B.G.C thank
the Swedish Research Council (VR) for financial support.


\expandafter\ifx\csname natexlab\endcsname\relax\global\long\def\natexlab#1{#1}
 \fi \expandafter\ifx\csname bibnamefont\endcsname\relax \global\long\def\bibnamefont#1{#1}
 \fi \expandafter\ifx\csname bibfnamefont\endcsname\relax \global\long\def\bibfnamefont#1{#1}
 \fi \expandafter\ifx\csname citenamefont\endcsname\relax \global\long\def\citenamefont#1{#1}
 \fi \expandafter\ifx\csname url\endcsname\relax \global\long\def\url#1{\texttt{#1}}
 \fi \expandafter\ifx\csname urlprefix\endcsname\relax\global\long\def\urlprefix{URL }
 \fi \providecommand{\bibinfo}[2]{#2} \providecommand{\eprint}[2][]{\url{#2}}

\end{document}